# High lattice thermal conductivity of hexagonal boron phosphide: A first principles study

[1]Rajmohan Muthaiah, [1]Fatema Tarannum, [2]N. Yedukondalu, [1]Jivtesh Garg

[1]School of Aerospace and Mechanical Engineering, University of Oklahoma, Norman, OK-73019, USA

[2]Department of Geosciences, Center for Materials by Design, and Institute for Advanced Computational Science, State University of New York, Stony Brook, New York 11794, USA

**Abstract:** Designing and searching for high lattice thermal conductivity materials in both bulk and nanoscale level is highly demanding for electronics cooling. Boron phosphide is a III-V compound semiconductor with superior structural and thermal properties. In this work, we studied lattice thermal conductivity of hexagonal boron phosphide (*h*-BP) using first principles calculations. For pure *h*-BP, we found a high lattice thermal conductivity (at 300K) of 561.2 Wm$^{-1}$K$^{-1}$ and 427.4 Wm$^{-1}$K$^{-1}$ along a-axis and c-axis respectively. These values are almost equal to hexagonal silicon carbide (2H-SiC) and cubic boron phosphide (c-BP). We also computed the length dependence thermal conductivity for its applications in nanostructures. At nanoscale (L=100 nm), a high thermal conductivity of ~71.5 Wm$^{-1}$K$^{-1}$(56.2 Wm$^{-1}$K$^{-1}$) is observed along a-axis (c-axis). This result suggests that, *h*-BP will be a promising material for thermal management applications in micro/nano electronics.

**Introduction:** High lattice thermal conductivity materials are of fundamental interest in improving thermal management in micro/nano electronics to improve system performance and reliability[1-3]. Boron based high thermal conductivity materials[4-10] are reported due to its light atomic mass and strong bonding between the constituent atoms. Thermal conductivity(*k*) of cubic[4, 8, 11] and monolayer[12, 13] boron phosphide was studied in detail. Nevertheless, *k* of its wurtzite structure is unknown. Hexagonal boron phosphide (*h*-BP) is a III-V compound semiconductor and only limited studies were carried out for absorption behavior (DFT-XCH approach)[14] and neutron detection[15]. In this work, by solving Peierls-Boltzmann Transport equation (PBTE) coupled with first principles calculations, we report a high thermal conductivity of ~561.2 Wm$^{-1}$K$^{-1}$(427.4 Wm$^{-1}$K$^{-1}$) for the pure bulk *h*-BP along a-axis (c-axis). We compared bulk and nanoscale lattice thermal conductivity of *h*-BP with hexagonal silicon carbide (2H-SiC), cubic boron phosphide(c-BP) and silicon (Si). Our first principles calculations reveal that, *k* of h-BP is almost equal to both 2H-SiC[16] and c-BP[4] as well as much higher than Si. Our results shed a pathway to explore *h*-BP for thermal management and other alliance applications.

**COMPUTATIONAL DETAILS:** All the first principles calculations were performed using plane-wave based QUANTUM ESPRESSO[17] package. We used local density approximation (LDA)[18] and norm-conserving pseudopotentials were used to represent core electrons. Self-consistence calculations were carried out with plane-wave energy cut-off of 90 Ry and Monkhorst-Pack[19] *k*-point mesh of 12 x 12 x 8. Calculations were carried out until the total energy is converged to $1e^{-12}$ Ry and the total force acting on each atom is $1e^{-5}$ eV/Å. Optimized *h*-BP with space group *P6$_3$mc* and lattice constants of a=3.14 Å and c/a=1.66 is shown in Fig 1. Dynamical matrix and harmonic (2$^{nd}$ order) interatomic force constants were computed on 8 x 8 x 6 q-grid. Anharmonic (3$^{rd}$ order) interatomic force constants were computed on 4 x 4 x 3 q-grid using QUANTUM ESPRESSO D3Q[20-22] package. Lattice thermal conductivity is calculated by solving Peierls-Boltzmann transport equation (PBTE)[20, 22, 23] iteratively within QUANTUM ESPRESSO thermal2 code with 30 x 30 x 15 **q**-mesh until the Δ*k* values are converged to $1.0e^{-5}$. Casimir scattering[24] is imposed for length dependence *k* calculations. Elastic constants and its related quantities were computed using QUANTUM ESPRESSO thermo_pw package using Voigt-Reuss-Hill (VRH) approximation[25].

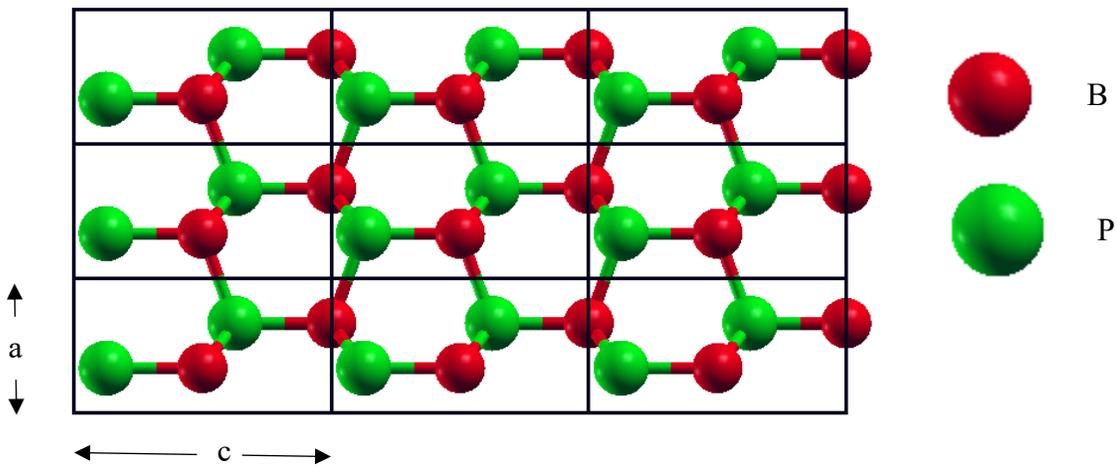

Figure 1: Atomic structure of hexagonal boron phosphide with lattice constants a=3.14 Å and c/a=1.66

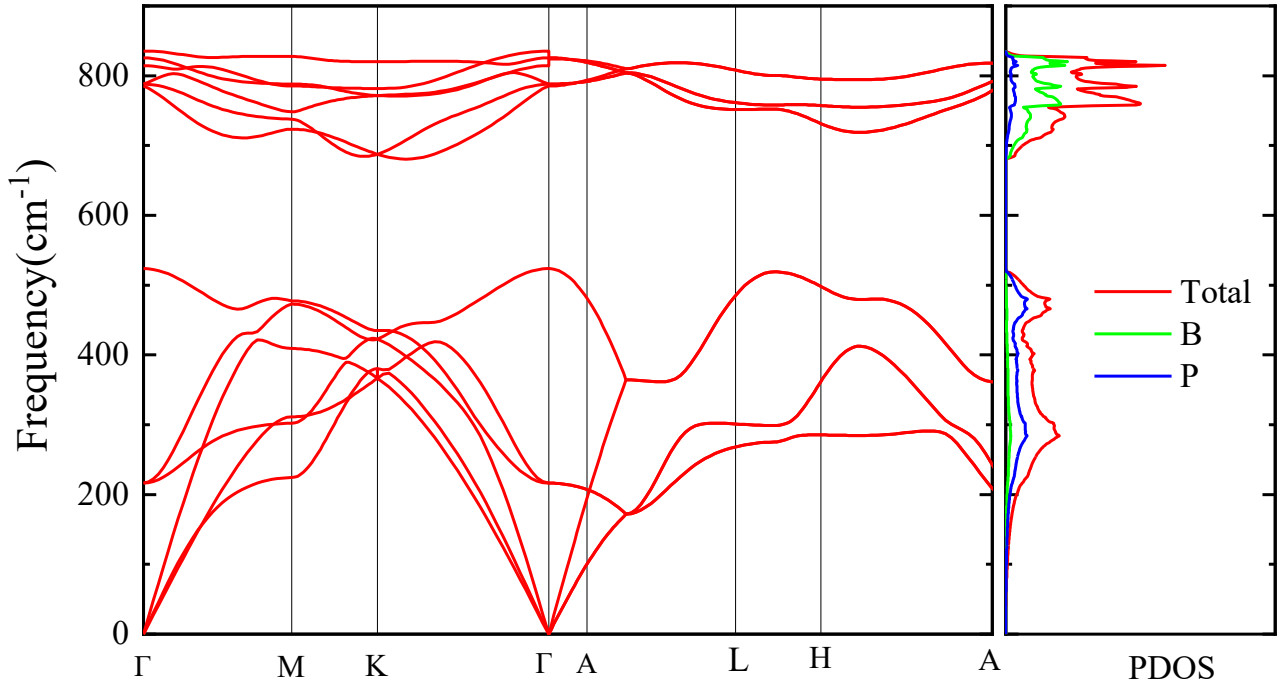

Figure 2: Phonon dispersion and phonon density of states for the *h*-BP

**Results and Discussion:** Calculated phonon dispersion and density of states(DOS) at equilibrium lattice constants (a = 3.14 Å and c/a = 1.66) is shown in Fig 2. We can observe phonons with high vibrational frequencies and are mainly attributed to the low atomic mass of boron (B) and phosphorous (P) as well as strong bonding between these atoms. For a material to be stable, it has to satisfy both mechanical and dynamical stability. To explore the mechanical stability, we computed the elastic constants (see in Table 1) and satisfied the well-known Born stability criteria[26] of $C_{44} > 0$, $C_{66} > 0$, $C_{66} = (C_{11}-C_{12})/2$, $C_{11} > C_{12}$ and $C_{33}(C_{11}+C_{12}) > 2(C_{13})$ indicates the system is mechanically stable. Phonons with no negative frequencies of all phonon modes indicate the dynamical stability of *h*-BP. Bulk modulus (B), Young modulus (E) and Shear modulus (G)

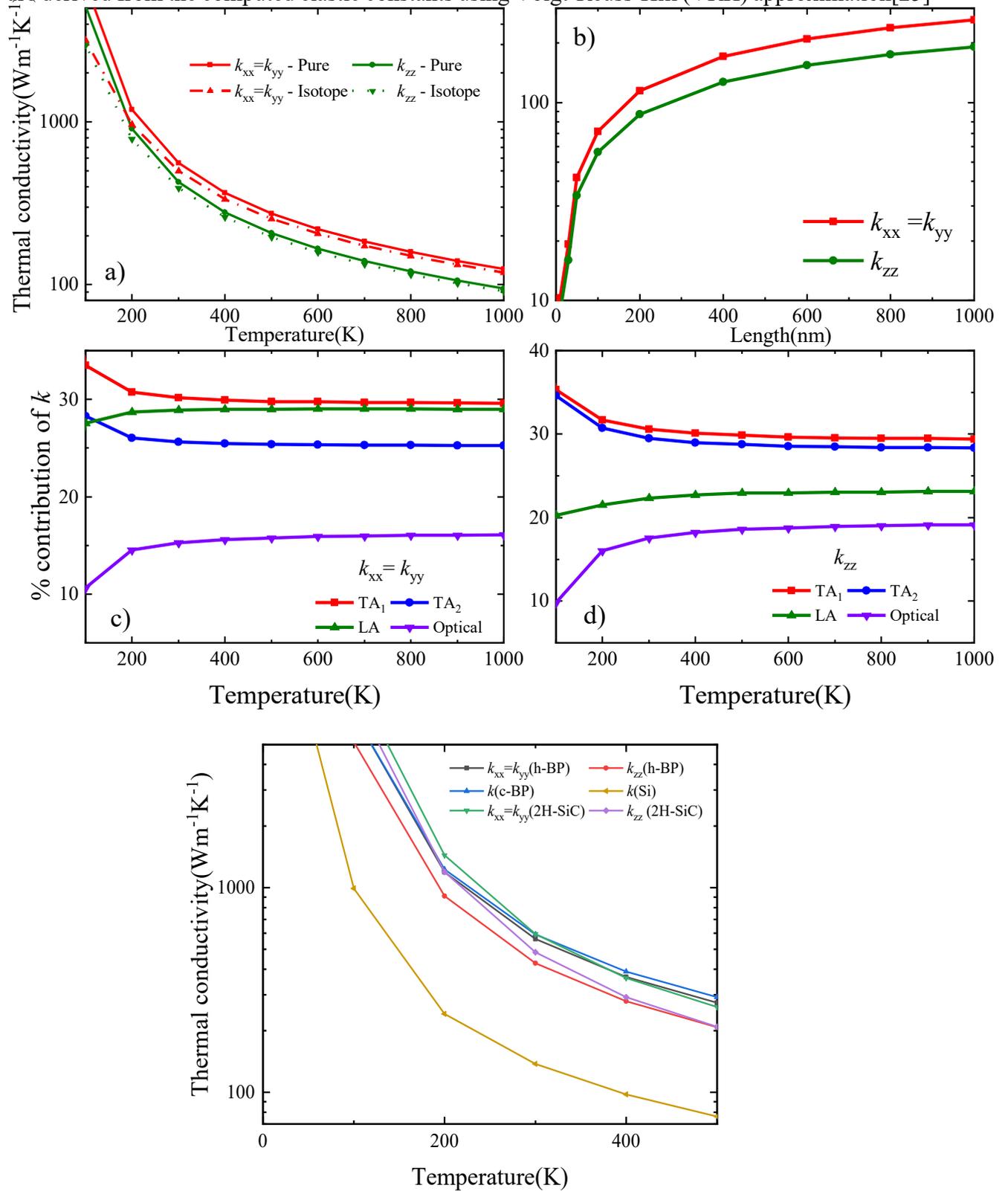

Figure 3a) Lattice thermal conductivity of isotopically pure and naturally occurring *h*-BP b) Length dependent lattice thermal conductivity at 300 K, c) and d) *k* contribution from transverse acoustic (TA), longitudinal acoustic (LA) and optical phonons along a- and c-axis e) comparison of $k_{h\text{-BP}}$ with silicon, c-BP and h-SiC

and are listed in Table 1. According to Pugh's ratio[27-29], B/G =0.984 <1.75, indicates the material is brittle in nature.

Table 1: Table 1: Elastic constants (in GPa) of h-BP, c-BP, 2H-SiC, h-BC2P h-BC6N and h-BC2N.

| Material | $C_{11}$ | $C_{33}$ | $C_{44}$ | $C_{66}$ | $C_{12}$ | $C_{13}$ | Bulk Modulus(B) | Young modulus(E) | Shear Modulus(G) |
|---|---|---|---|---|---|---|---|---|---|
| h-BP | 440.63 | 491.89 | 144.10 | 191.37 | 57.88 | 21.22 | 174.85 | 398.23 | 177.73 |
| c-BP | 363.88 |  | 203.58 |  | 81.81 |  | 175.83 | 395.45 | 175.77 |
| 2H-SiC | 522.5 | 557.73 | 156.77 | 214.95 | 92.63 | 43.61 | 218.83 | 453.21 | 196.44 |
| h-BC6N[7] | 1182.98 | 1298.11 | 438.90 | 537.20 | 108.58 | 20.33 | 440.00 | 1107.40 | 512.30 |
| h-BC2P[5] | 675.00 | 680.60 | 198.00 | 305.00 | 65.00 | 30.80 | 253.60 | 582.20 | 260.60 |
| h-BC2N[30] | 1049.00 | 1054.52 | 406.97 | 468.39 | 112.32 | 65.18 | 404.00 | 985.00 | 450.00 |

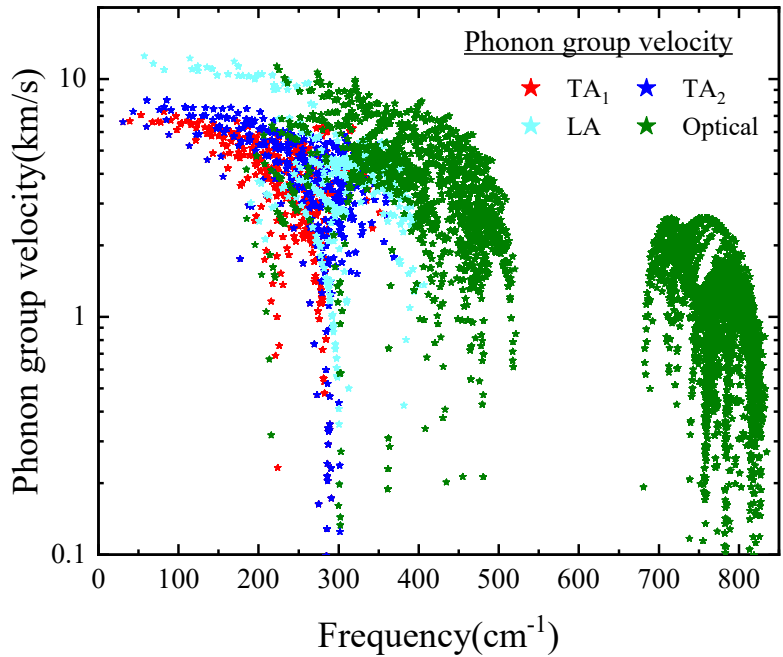

Figure 4: Phonon group velocities of TA$_1$, TA$_2$, LA and optical phonons for *h*-BP

We computed the temperature dependent lattice thermal conductivity(*k*) of both isotopically pure and isotopically scattered *h*-BP on its relaxed structure along a-axis and c-axis and is shown in Fig. 3a. At room temperature(300 K), *k* for an isotopically pure (naturally occurring) *h*-BP is 561 Wm$^{-1}$K$^{-1}$(500 Wm$^{-1}$K$^{-1}$) and 427 Wm$^{-1}$K$^{-1}$(392 Wm$^{-1}$K$^{-1}$) along a-axis and c-axis, respectively. This ~11% (8.8%) reduction in *k* due to isotope scattering along a-axis(c-axis) is due to the mass variation in isotopes of 19.9 % of $^{10.012}$B and 80.1% $^{11.009}$B. There is no mass variation in phosphorous (P) [31] and hence the difference is low. *k* of *h*-BP in comparison with related materials such as c-BP, *h*-Si is shown in Fig 3e. At 300 K, *k* of *h*-BP is almost equal to the *c*-BP and 2H-SiC and ~300% higher than Si.

To understand this high thermal conductivity, we investigated the phonon group velocities, phonon lifetime (inverse of phonon linewidth) of transverse acoustic (TA$_1$ and TA$_2$) and longitudinal acoustic (LA) as well as optical phonons. We also computed the *k* contribution from TA$_1$, TA$_2$, LA and optical phonon modes and is shown in Fig 3c and d. At 300 K, TA$_1$, TA$_2$, LA and optical phonons contributes to 30.18% (30.60%), 25.65% (29.51%), 28.82% (22.35%) and 15.35% (17.54%) respectively along a-axis(c-axis). Interestingly all acoustic phonons have almost equal contribution. LA phonons with maximum phonon group velocity (> 10 km/s) combined with moderate scattering rate contributes to 28.82% in thermal conductivity. On the thermal hand, TA phonons which has lower phonon group velocities than LA phonons but with lower scattering rate causing to contribute to ~25-30% in *k*. This high thermal conductivity is due to high phonon group velocities ($v_{\alpha\lambda} = \partial\omega_\lambda/\partial q$) and high phonon lifetime (inverse of phonon linewidth) as shown in Fig 4. and Fig 5. respectively. High phonon frequency and phonon group velocities are mainly attributed with light atomic mass and high bonding strength between the atoms which can be clearly observed from the calculated elastic (stiffness) constants (see Table 1). Surprisingly, *h*-BP has almost equal B, G and E of the c-BP but are lower than 2H-SiC, $BC_6N$ and $BC_2N$ etc.

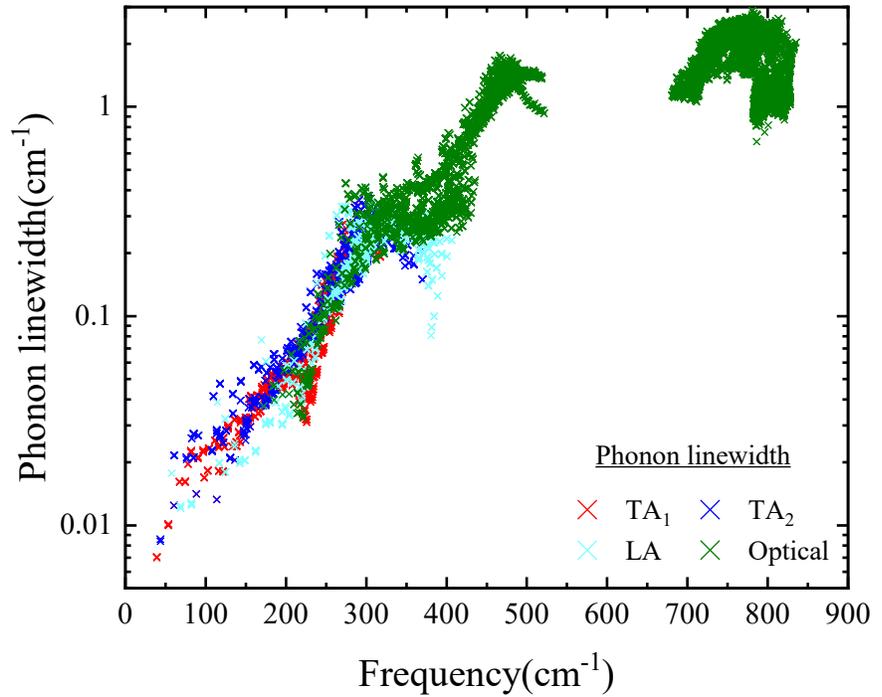

Figure 5: Phonon linewidth of TA$_1$, TA$_2$, LA and optical phonons for *h*-BP

Since *k* in nanostructured materials is critical, we also computed the length dependent lattice thermal conductivity between 10 nm to 1000 nm to explore the thermal transport in nanostructures and is shown in Fig 3b. At 300 K, for L= 100 nm, *k* of *h*-BP is 71.44 $Wm^{-1}K^{-1}$ along a-axis which is significantly higher than the thermal conductivity of $k_{L=100}$ nm of silicon[6]. This is mainly attributed with contribution of all phonon modes ($TA_1$, $TA_2$, LA and optical phonon modes) with phonon mean free path less than 100 nm as shown in Fig 6. At $\omega > 250$ $cm^{-1}$, we can observe a significant contribution from both acoustic and optical phonons which has a phonon mean-free-path of <100 nm, contributing to the high nanoscale thermal conductivity in *h*-BP.

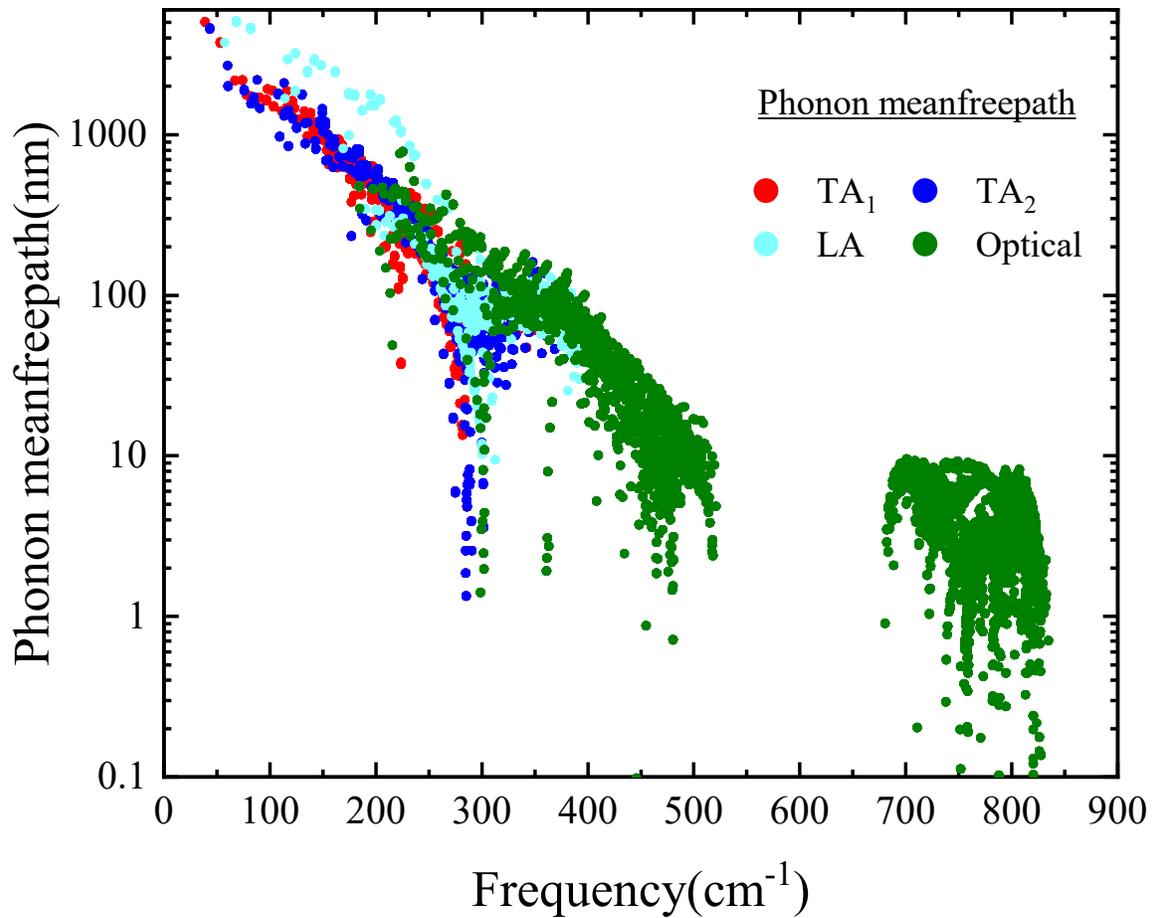

Figure 6: Phonon mean-free path of $TA_1$, $TA_2$, LA and optical phonons for hexagonal boron phosphide

**Conclusions:** To summarize, this work report a high lattice thermal conductivity of 561 Wm$^{-1}$K$^{-1}$(427 Wm$^{-1}$K$^{-1}$) along a-axis(c-axis) for pure bulk *h*-BP using first principles calculations and compared it with c-BP, Si and *h*-SiC and our calculations shows that *h*-BP has 300% higher *k* than Si and almost equal to c-BP and *h*-SiC. We also report a high nanoscale (L=100 nm) thermal conductivity (71.44 Wm$^{-1}$K$^{-1}$) arising from contribution from all phonon modes ω > 250 cm$^{-1}$ with phonon mean free path in the range of 100 nm. This high thermal conductivity is due to high phonon frequencies and phonon group velocities arising from the strong bond between boron and phosphorous (B-P) and the light atomic mass of the constituent atoms B and P. Surprisingly, bulk modulus, elastic modulus and shear modulus of *h*-BP are almost equal to the c-BP. This high thermal conductivity in both bulk and nanoscale points to *h*-BP as a potential candidate for nanoscale thermal management applications.

**Conflicts of Interest**

There are no conflicts of interest to declare.

**Acknowledgements**

RM, FT and JG acknowledge support from National Science Foundation CAREER award under Award No. #1847129. We also acknowledge OU Supercomputing Center for Education and Research (OSCER) for providing computing resources for this work.